\documentclass[]{aa}

\usepackage[utf8]{inputenc}
\usepackage{graphicx}
\usepackage[varg]{txfonts}

\bibpunct{(}{)}{;}{a}{}{,}

\begin{document}
\title{Extended neutral hydrogen filamentary network in NGC 2403}

\author{Simone Veronese\inst{\ref{astron},\ref{kapteyn}} \and W. J. G. de Blok\inst{\ref{astron},\ref{kapteyn},\ref{uct}} \and F. Walter\inst{\ref{mpia},\ref{nrao}}}

\institute{Netherlands Institute for Radio Astronomy (ASTRON), Postbus 2, 7990 AA Dwingeloo, The Netherlands, \email{veronese@astron.nl}\label{astron}
\and Kapteyn Astronomical Institute, University of Groningen, PO Box 800, 9700 AV Groningen, The Netherlands\label{kapteyn}
\and Department of Astronomy, University of Cape Town, Private Bag X3, 7701 Rondebosch, South Africa\label{uct}
\and Max Planck Institute for Astronomy, Königstuhl 17, D-69117 Heidelberg, Germany\label{mpia}
\and National Radio Astronomy Observatory, Pete V. Domenici Array Science Center, P.O. Box O, Socorro, NM 87801, USA\label{nrao}}

\date{Received <date> / Accepted <date>}

\abstract{We present new neutral hydrogen (H\,{\sc i}) observations of the nearby galaxy NGC 2403 to determine the nature of a low-column density cloud that was detected earlier by the Green Bank Telescope.\\We find that this cloud is the tip of a complex of filaments of extraplanar gas that is coincident with the thin disk. The total H\,{\sc i} mass of the complex is $2\times10^{7}\text{ M}_\odot$ or 0.6\% of the total H\,{\sc i} mass of the galaxy. The main structure, previously referred to as the 8-kpc filament, is now seen to be even more extended, along a 20 kpc stream.\\The kinematics and morphological properties of the filaments are unlikely to be the result of outflows related to galactic fountains. It is more likely that the 20 kpc filament is related to a recent galaxy interaction. In this context, a $\sim$ 50 kpc long stellar stream has been recently detected connecting NGC 2403 with the nearby dwarf satellite DDO 44. Intriguingly, the southern tip of this stream overlaps with that of 20 kpc H\,{\sc i} filament.\\We conclude that the H\,{\sc i} anomalies in NGC 2403 are the result of a recent ($\sim2\text{ Gyr}$) interaction with DDO 44 leading to the observed filamentary complex.}

\maketitle

\section{Introduction}\label{sec:intro}
Galaxies form stars through the collapsing of giant molecular (mainly molecular hydrogen, H\,{\sc ii}) clouds on timescales of $\sim10^7$ yr \citep{meidt15,schinnerer19,walter20} and over the cosmic time \citep{madau14} galaxies progressively deplete their molecular gas content. In a perfect steady state, the H\,{\sc ii} reservoir is replenished by the cooling of the atomic hydrogen (H\,{\sc i}) in the interstellar medium \citep{clark12,walch15}, which will cause a reduction in the content of the atomic gas. However, both simulations and observations reveal that the H\,{\sc i} content in galaxies is almost constant from $z\sim1$ \citep{dave17,chen21}. Consequently, in order to maintain star formation over cosmic time, galaxies must accrete H\,{\sc i}.\\Previous studies of H\,{\sc i} interaction features and dwarf galaxies have shown that they do not provide enough cold gas to replenish the material reservoir for star formation \citep{sancisi08,putman12,blok20}. Other extraplanar gas observed closer to the galaxy disks is usually related to galactic fountains \citep{putman12,li21,marasco22}: star formation and supernova explosions eject interstellar medium from the disk to scale-heights of $\sim$ kpc. The gas catches material from the circum-galactic medium and falls back onto the galaxy carrying new gas to fuel the star formation. However, the amount of material accreted with this process is again not enough to replenish the atomic gas reservoir \citep{putman12,afruni21}.\\A hypothesis that could explain the constancy of the H\,{\sc i} content is that galaxies accrete gas from the Inter-Galactic Medium (IGM) \citep{sommerville15,danovich15}. The mode of this accretion, i.e., cold clouds and streams or hot diffuse gas \citep{dekel06,danovich15}, is uncertain as well as the accretion rate. Nevertheless, this model could indicate a possible way to solve the problem of how to sustain the star formation activities over cosmic time.\\It has, however, one major limitation: the directly detectable component of this gas, namely the warm H\,{\sc i} at $\sim10^{4}\text{ K}$, forms only a small fraction ($<1\%$) of the total amount of accreting material which mostly consists of warm-hot 10\textsuperscript{5} K gas. This leads to low H\,{\sc i} column densities of $\sim10^{17}\text{ cm}^{-2}$ \citep{popping09}. This value is at least one order of magnitude lower than the detection limits of the current radio interferometer observations. Single-dish telescopes can reach these column densities but only at low resolutions that usually do not resolve the relevant scales. So far, surveys that could potentially detect accreting H\,{\sc i} clouds have not found any direct evidence of them \citep{hipass,blok02,pisano04,alfalfa,things,halogas}.\\In this paper, we concentrate on one potential low-column density cloud found with the Green Bank Telescope (GBT) near the well-studied spiral galaxy NGC 2403 \citep{blok14}. The cloud is located $\sim16$ kpc ($\sim2R_{25}$) to the NW of the centre of the galaxy but the low spatial resolution of the data does not allow to properly distinguish it from the disk emission and to robustly constrain its properties. The motivation behind this study is to confirm its detection with high spatial resolution radio interferometric data and to understand its nature, especially, if it is indeed the signature of IGM accretion that previous studies have tried to find. With high-resolution data we can also test other scenarios, for example whether it could be a result of strong outflow, or perhaps the remnant of a galactic interaction, and how it is linked to the kinematically anomalous H\,{\sc i} features first detected by \citet{fraternali01} (see also \citealt{fraternali02,fraternali06,things,blok14}).\\NGC 2403 is a member of the M81 group, located at a distance of 3.18 Mpc\citep{madore91} ($1'=0.93$ kpc, $1''=0.015$ kpc) with a systemic velocity of 133.2 km\,s$^{-1}$ \citep{fraternali02}. The main H\,{\sc i} disk is inclined by 63° \citep{fraternali02} and has a mass of $3.24\times10^{9}\text{ M}_\odot$ \citep{fraternali02}. A list of the main optical and radio parameters can be found in \citet{fraternali02}.\\We obtained new H\,{\sc i} radio observations of NGC 2403 and the GBT cloud in order to determine the nature of this cloud and explore its connection with the previously reported 8-kpc anomalous velocity filament \citep{fraternali02}.\\In Sec. \ref{sec:red} we describe the data reduction and the creation of the data products that were analysed following the methods illustrated in Sec. \ref{sec:ana}. The main results and interpretation of our study are given in Sec. \ref{sec:disk} and are summarised in Sec. \ref{sec:conc}.

\section{Observations}
The new Very Large Array (VLA) observations discussed in this paper were taken between December 5th 2019 and December 9th 2019 with the Expanded VLA in D-configuration\footnote{Project ID: 19B-081.} (hereafter referred to as the `2019 data'). The L-band bandwidth of 5.3 MHz comprised 2700 1.9 kHz (0.4 km\,s$^{-1}$) channels with a central frequency of 1.41701 GHz. Two pointings were observed: 3-hours centred on the galaxy ($\alpha=7\textsuperscript{h}36\textsuperscript{m}44\textsuperscript{s}$, $\delta=65^{\circ}35'35''$) and another 3-hours centred on the GBT cloud ($\alpha=7\textsuperscript{h}35\textsuperscript{m}00\textsuperscript{s}$, $\delta=65^{\circ}44'4''$).\\In addition to these new data we also used archival VLA H\,{\sc i} data of NGC 2403 centred on the galaxy. The archival data comprise The H\,{\sc i} Nearby Galaxy Survey (THINGS) observations\footnote{We included the data from the C- and D-array configurations only.} \citep{things} and those discussed in \citet{fraternali01}. Both data sets have a velocity resolution of 5.2 km\,s$^{-1}$.

\subsection{Data Reduction}\label{sec:red}
Here we give a brief description of the procedure used to reduce the new 2019 dataset. The data reduction was carried out with the software CASA v5.8.0 \citep{casa}. A preliminary data flagging was performed on the primary and secondary calibrator data (to correct for shadowing, correlator glitches, misbehaving antennas, bandwidth cut-offs, Radio Frequency Interference (RFI), Milky Way emission and absorption) as well as on NGC 2403 (to take into account shadowing, correlator glitches, misbehaving antennas) with the CASA task \textit{flagdata}. The RFI was manually removed in CASA \textit{plotms}. A calibration model was then built with the following corrections: target elevation, delay, bandpass, complex gain and flux scale using the CASA tasks \textit{gencal}, \textit{gaincal} and \textit{fluxscale}. A set of weights for the data was constructed based on the variance of the data with the CASA task \textit{statwt} and a Hanning-smoothing in frequency was applied to remove Gibbs ringing artefacts with the CASA task \textit{hanningsmooth}. Finally, the continuum was estimated with a simple linear fit, justified by the narrow bandwidth, to the amplitudes and phases in the line free channels and subtracted from the data with the CASA task \textit{uvcontsub}.\\The THINGS and the 1999 data were re-reduced using similar standard procedures.

\subsection{Cube creation}
We used the CASA task \textit{tclean} to create a combined mosaicked H\,{\sc i} cube of the two pointings of the 2019 data. We applied a robustness parameter of 0.5 and a tapering of $48''$ to the data in order to balance spatial resolution and sensitivity. Choosing a pixel size of $9''$ to properly sample the beam, and accounting for the primary beam size then yielded channel maps with the extent of $400\times400$ pixels. As our aim is to combine the new observations with the archival data, we chose the channel width to be 5.3 km\,s$^{-1}$, leading to 209 channels covering the range $-377.8$ to $724.6\text{ km\,s$^{-1}$}$.\\\textit{tclean} was run in the \textit{auto-multithresh} masking mode: for each iteration of the cleaning process and for each channel of the cube a mask that enclosed the emission above $4\times\sigma$ in the channel is created. Inside the masked region \textit{tclean} smooths the data by a factor of 1.5 and cleans until it reaches the stopping threshold of 0.5$\sigma$, where the noise is now the one of the smoothed channel. This results in a cube based on only the new data.\\In addition, we use \textit{tclean} to combine all the available observations including the archival data into a single H\,{\sc i} cube and use the same imaging parameters as the mosaicked 2019 cube above. As the different uv distributions in the two pointings would lead to different resolutions over the mosaic, we applied a $48''$ taper which resulted in a common beam size for the entire mosaic. Furthermore, the noise is different for the two pointings, but at every position the mosaic has the highest signal to noise possible given the data. We will return to this below. The more limited velocity range of the archival data resulted in a cube with 102 channels of 5.3 km\,s$^{-1}$.

\section{Data analysis}\label{sec:ana}
\subsection{Cube statistics}
Throughout this paper, the given noise values are referring to the non-primary beam corrected cubes, while all the flux density and H\,{\sc i} mass measurements have been performed on primary beam corrected data.\\The 2019 H\,{\sc i} cube has a beam size of $62\times55\text{ arcsec}$ ($1.24\times1.1\text{ kpc}$) and the single-channel median noise is $\sigma=0.53\text{ mJy\,beam$^{-1}$}$. This can be converted into an H\,{\sc i} column density with the equation
\begin{equation}
    N_{HI}=1.25\times10^{24}\frac{F\Delta v}{A} \text{ cm}^{-2}
\end{equation}
where $F$ is the flux density in Jy\,beam$^{-1}$, $\Delta v$ is the velocity width in km\,s$^{-1}$, and $A$ is the beam area in arcsec\textsuperscript{2} calculated assuming a Gaussian beam: $A=1.13ab$, where $a$ and $b$ are the beam major and minor axis in arcsec, respectively. The resulting 3$\sigma$ 1-channel column density is $2.7\times10^{18}\text{ cm}^{-2}$.\\The H\,{\sc i} cube which also includes the archival data has a beam size of $52\times49\text{ arcsec}$ ($1.04\times0.98\text{ kpc}$). The noise in this cube is not spatially constant due to the combination of different pointings with different integration times. The median $\sigma$ in the NW pointing is 0.31 mJy\,beam$^{-1}$, while in the central pointing it is 0.23 mJy\,beam$^{-1}$. However, this difference is not relevant, because in the analysis we opt to use the global noise value of $\sigma=0.295$ mJy\,beam$^{-1}$, giving a 3$\sigma$ 1-channel H\,{\sc i} column density of $2\times10^{18}\text{ cm}^{-2}$, as an indicative sensitivity for the combined cube.

\begin{figure*}
    \resizebox{\hsize}{!}
    {\includegraphics[]{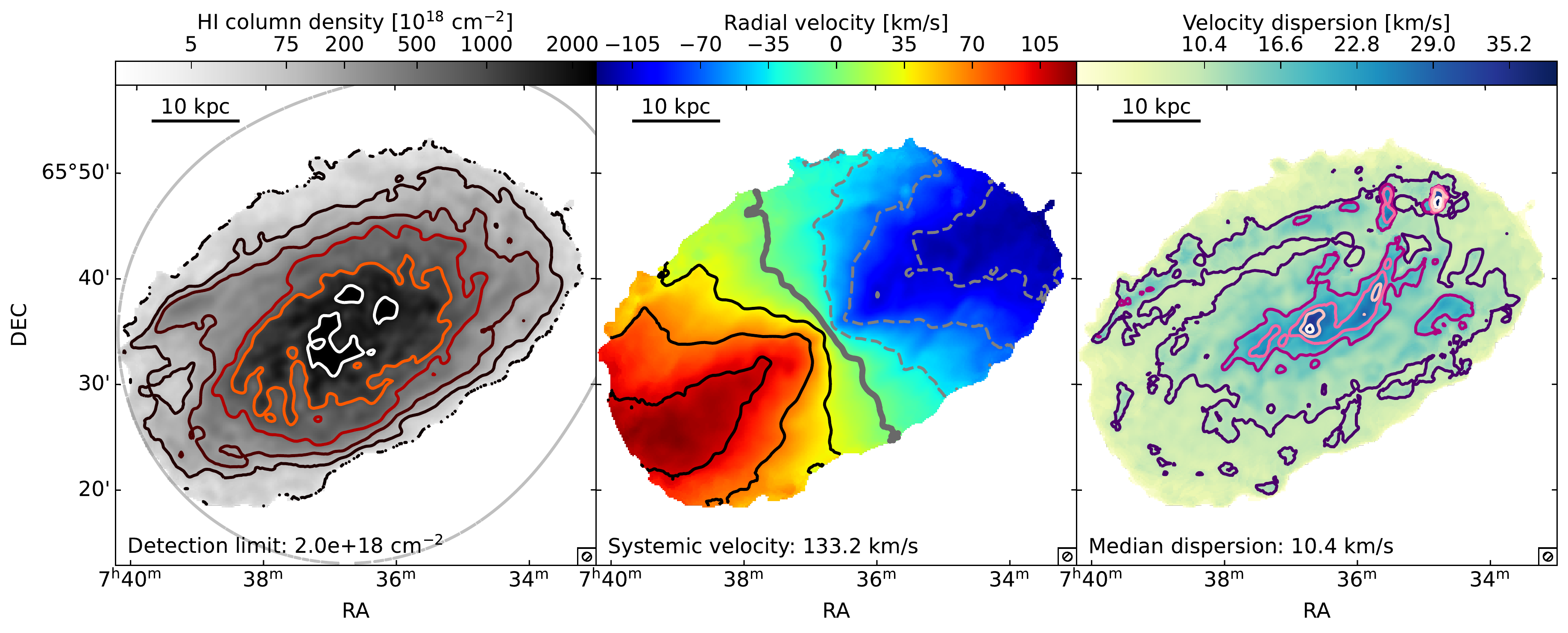}}
    \caption{\textit{Left panel}: primary beam corrected H\,{\sc i} column density map of NGC 2403 with reversed grey-scale colormap from faint (grey) to bright (black) emission. The sharp cut-off in the bottom left is due to the 20\% cut-off threshold of the primary beam response, which is illustrated with the light grey curve. Contours levels are from 5 $\times$ 10\textsuperscript{18} cm\textsuperscript{-2} to 2 $\times$ 10\textsuperscript{21} cm\textsuperscript{-2}. The column density limit $3\times\sigma\times dv=2\times10^{18}$ cm\textsuperscript{-2}, where $dv=5.3$ km\,s$^{-1}$ is the spectral resolution, is reported at the bottom. In the bottom-right is shown the $52''\times49''$ beam. \textit{Central panel}: intensity-weighted mean velocity field of NGC 2403 with respect to the systemic velocity of 133.2 km\,s$^{-1}$, given at the bottom. Dashed contours ($-$105, $-$70, $-$35) km\,s$^{-1}$ refer to the approaching side, while solid lines correspond to radial velocities of (35, 70, 105) km\,s$^{-1}$. The thick grey line is the kinematical minor axis, where the line-of-sight velocity is 0 km\,s$^{-1}$. \textit{Right panel}: second moment map, where the signature of the filaments with their anomalous velocities is visible in the increased second-moment values.}
    \label{fig:maps}
\end{figure*}
\begin{figure*}
    \resizebox{\hsize}{!}
    {\includegraphics[]{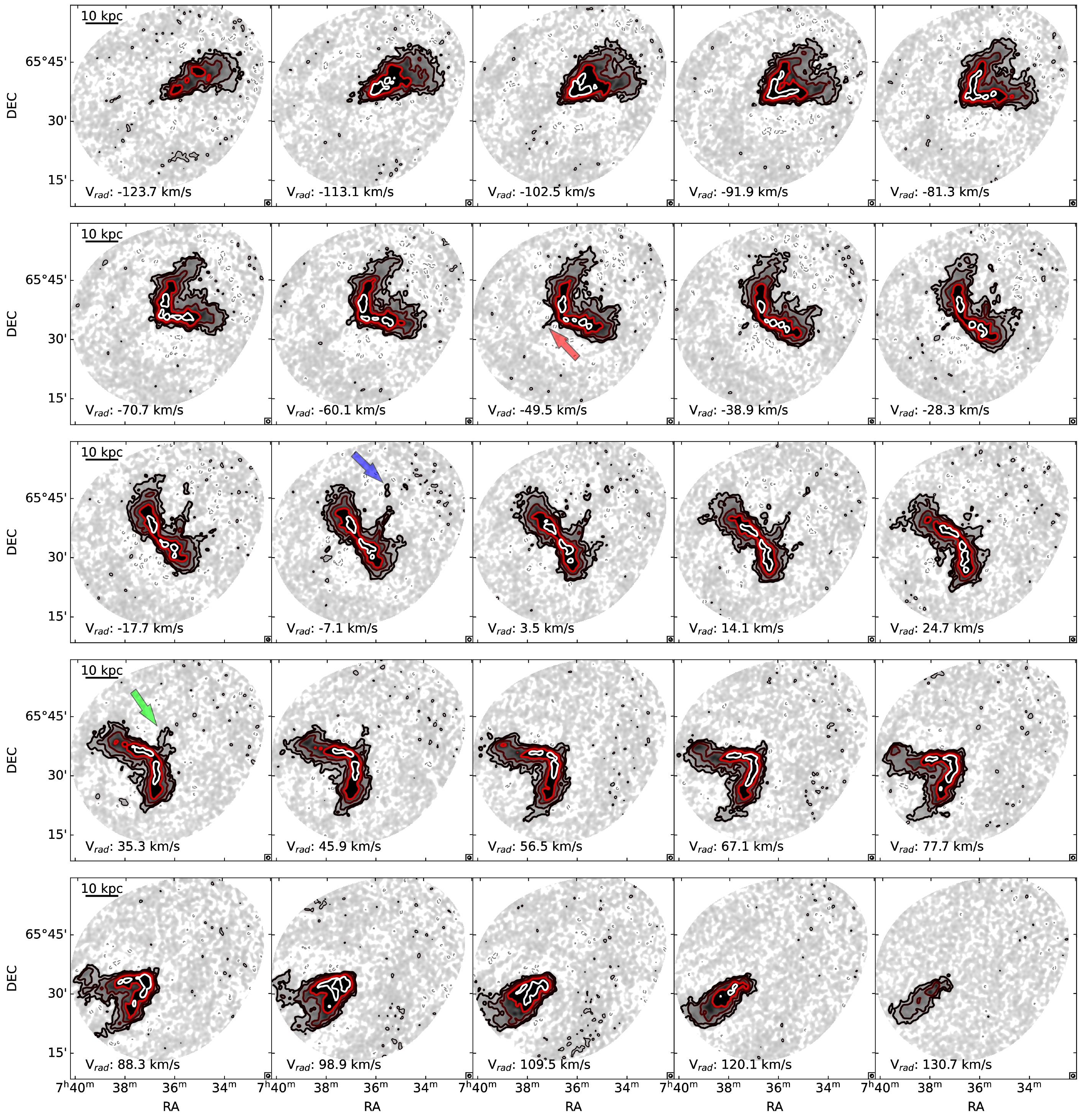}}
    \caption{Channel maps of the data cube with both new and archival observations where we show only every second channel. The colour-scale contours correspond to 2 $\times$ (1, 4, 16, 64) $\times$ 10\textsuperscript{18} cm\textsuperscript{-2}, where $2\times10^{18}$ cm\textsuperscript{-2} is the $3\sigma$ 1-channel column density limit. The thick contour levels indicate the significant emission identified by the SoFiA source finder. Most of the low-level structures shown in thin contours are noise or mosaicking artefacts. The dashed grey contours denote the column density of $-$2 $\times$ 10\textsuperscript{18} cm\textsuperscript{-2}. The egg-shape borders are due to the 20\% cut-off threshold of the primary beam response. At the bottom we report the line-of-sight velocity (w.r.t. the systemic velocity 133.2 km\,s$^{-1}$) in each channel. The filamentary detections, labelled with coloured arrows, are evident between $-$59 km\,s$^{-1}$ and 47 km\,s$^{-1}$.}
    \label{fig:chanmap}
\end{figure*}
\begin{figure*}
    \resizebox{\hsize}{!}
    {\includegraphics[]{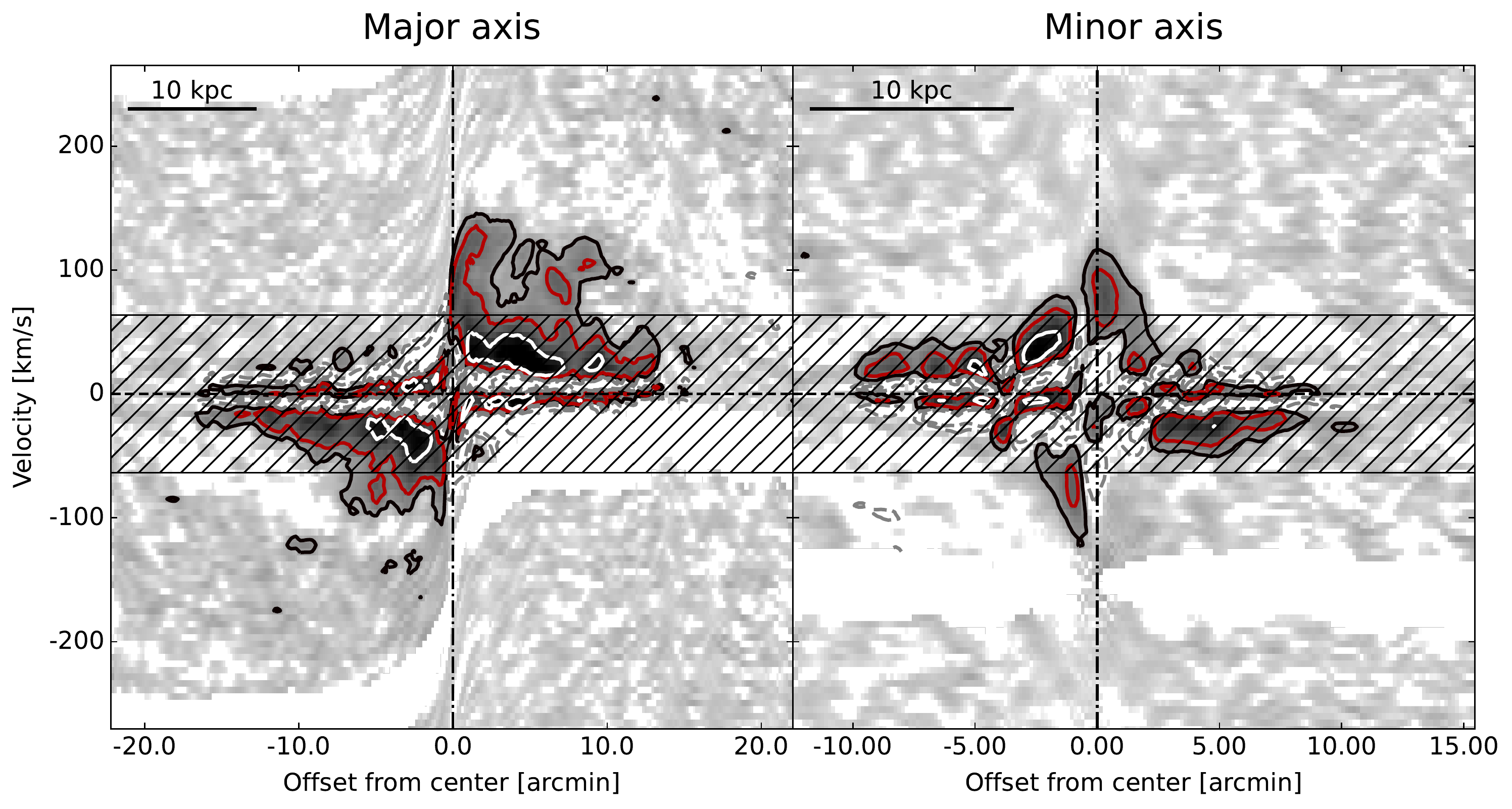}}
    \caption{Position-velocity diagram along the major axis (left panel) and minor axis (right panel) through a beam-wide slice of the shuffled data cube without the thin disk removal. The emission is mapped with a reversed grey-scale and we overlaid the 2 $\times$ (1, 3, 9) $\times10^{18}$ cm\textsuperscript{-2} levels with grey-scale contours. The blanked area in the top-left and blanked stripe in the bottom are the result of the Milky Way filtering. The grey dashed lines denote the column density of $-3\sigma$. The channels between $\pm$ 65 km\,s$^{-1}$ with the residual thin disk and diffuse extraplanar emission are highlighted with the hatched area. They were not used for the computation of the moment map shown in the central panel of Fig. \ref{fig:filmap}.}
    \label{fig:shufflepv}
\end{figure*}
\begin{figure*}
    \resizebox{\hsize}{!}
    {\includegraphics[]{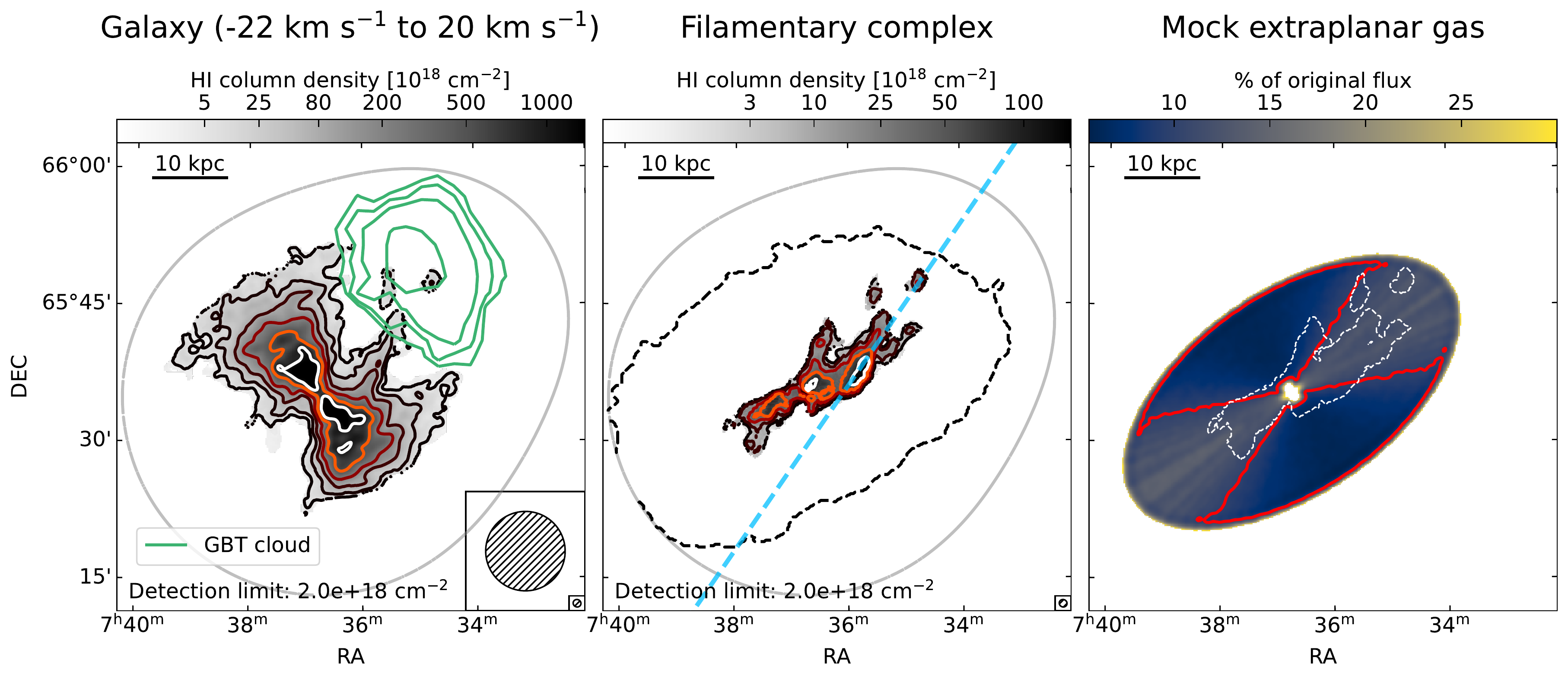}}
    \caption{\textit{Left panel}: primary beam corrected H\,{\sc i} column density map of NGC 2403 (in reversed grey-scale colormap) overlaid with the GBT cloud candidate (green) from \citet{blok14}. The map was produced by integrating over the channels from $-22$ km\,s$^{-1}$ to 20 km\,s$^{-1}$ w.r.t. the systemic velocity. This range corresponds to the one used by \citet{blok14} to compute the candidate GBT cloud moment 0 map. Grey-scale contour levels are from 5 $\times$ 10\textsuperscript{18} cm\textsuperscript{-2} to 1 $\times$ 10\textsuperscript{21} cm\textsuperscript{-2}. The $3\sigma$ 1-channel column density limit is reported at the bottom. The beam of the interferometric data for both the GBT and our VLA observations is shown in the bottom right. The green contours, instead, define the (6.25, 12.5, 25, 62.5) × 10\textsuperscript{17} cm\textsuperscript{-2} column density in the GBT data. The light-grey line denotes the 20\% cut-off threshold of the primary beam response. \textit{Central panel}: primary beam corrected H\,{\sc i} column density map of the anomalous-velocity gas. Grey-scale contour levels are (3, 10, 25, 50, 100) $\times$ 10\textsuperscript{18} cm\textsuperscript{-2} column density. The spatial scale is the same as in the left panel, as well as the definition of the light grey line. We indicate the galaxy edge with the dashed-black contour. The light blue dashed line represents the path along which the position-velocity slice has been extracted (see Fig. \ref{fig:filpv}). \textit{Right panel}: Illustrative example of the selection effect introduced by the shuffle procedure (see discussion in section \ref{sec:disk}). The map shows the fraction of the original extraplanar flux density of a mock galaxy we were able to recover and the red contour encloses the region where we are severely affected by selection effect, i.e., where more than 90\% of the original flux density is lost. The blanked bits in the centre denote the region where the result is unreliable because of the steep gradient in velocity. The dashed white contours are, instead, the region occupied by the observed filamentary complex.}
    \label{fig:filmap}
\end{figure*}
\begin{figure*}
    \resizebox{\hsize}{!}
    {\includegraphics[]{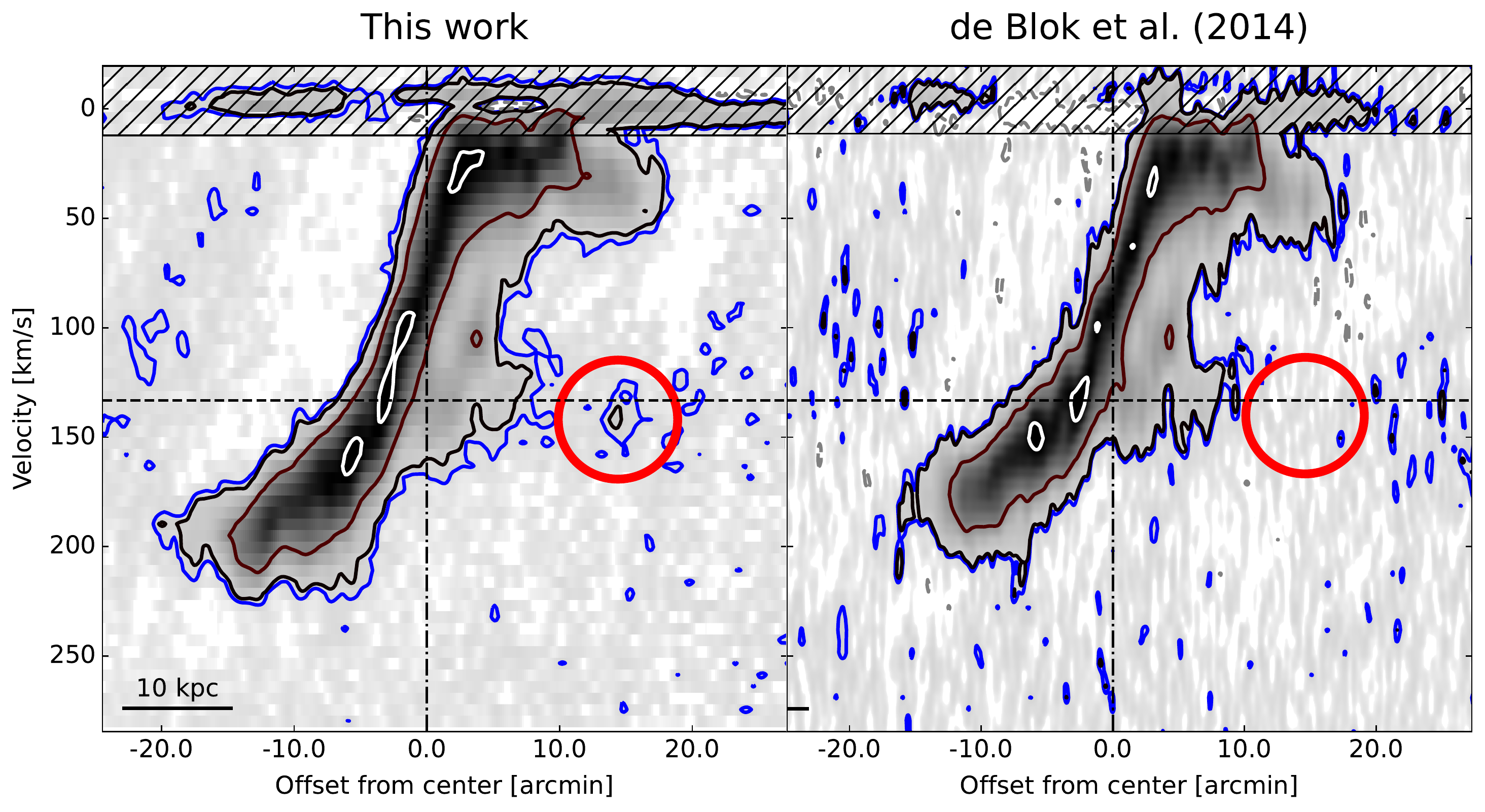}}
    \caption{Comparison between the position-velocity diagram through the centres of the cloud candidate and the 20 kpc filament along a 100''-thick slice as taken through our data (left panel) and as shown in Fig. 6 in \citet{blok14} (right panel). The emission is mapped with a reversed grey-scale and we overlaid the 2.5 $\times$ (1, 9, 81) $\times$ 10\textsuperscript{18} cm\textsuperscript{-2} column density levels with colour-scale contours (black, red and white, respectively). The blue contour denotes a level of 1.2 $\times$ 10\textsuperscript{18} cm\textsuperscript{-2} and highlights the detection of the 20 kpc filament tip with the combination of the new and archival VLA data. The red circles highlight the tip of the 20 kpc filament which was not detected in the \citet{fraternali02} and \citet{blok14} analysis.}
    \label{fig:filpv}
\end{figure*}
\begin{figure*}
    \resizebox{\hsize}{!}
    {\includegraphics[]{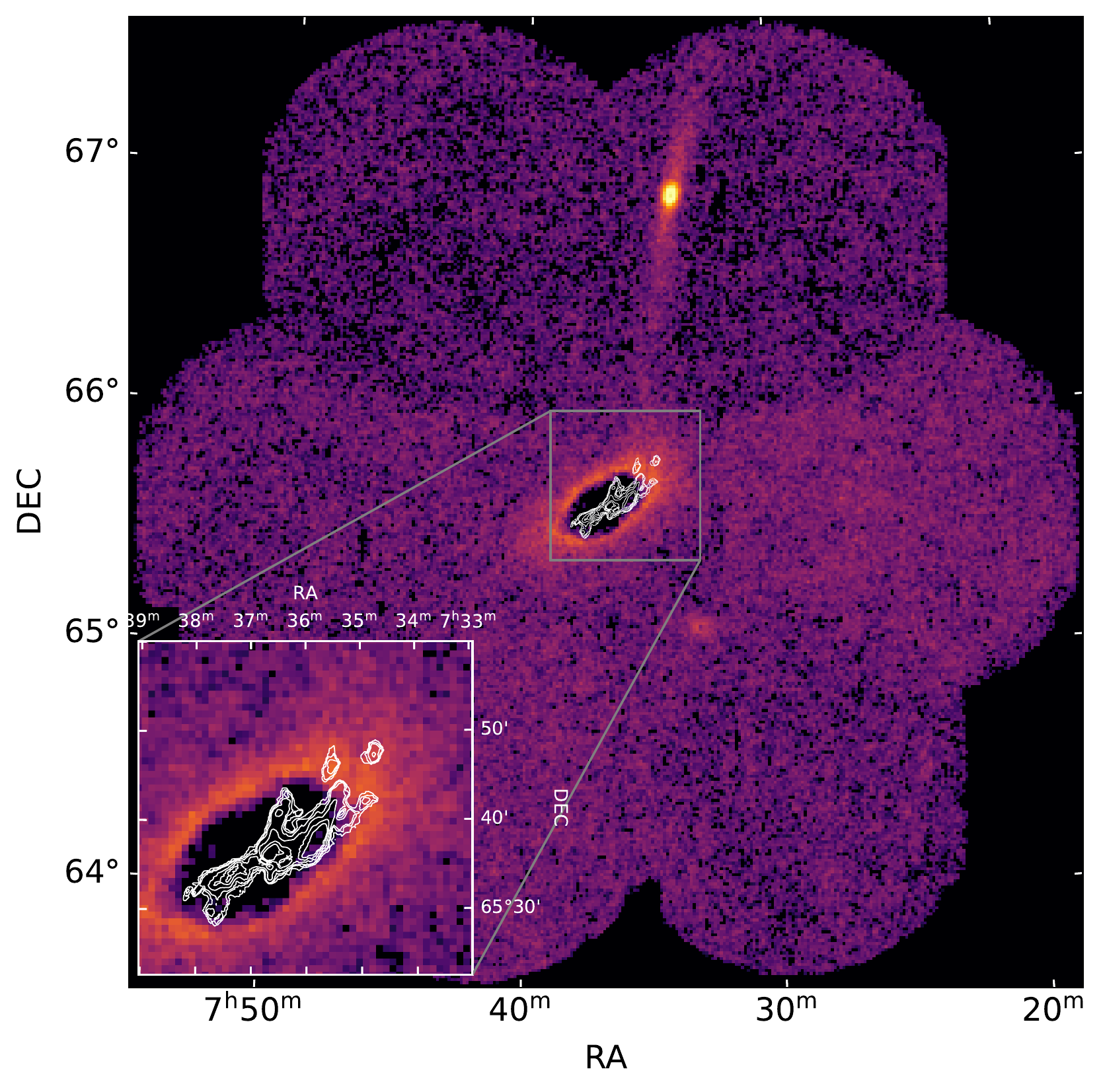}}
    \caption{Overlay between the re-reduced density map of candidate RGB stars presented in \citet{carlin19} and the H\,{\sc i} filaments observed in our data (white contours). Bins in the density map are completeness-corrected and have a resolution of 0.75 arcmin. The hole in the centre is due to extreme crowding. The colours denote the number of stars per 0.75' bin. The stellar stream is connecting DDO 44 and NGC 2403 with the tip of the H\,{\sc i} filaments, strongly suggesting that DDO 44 and NGC 2403 underwent a recent interaction.}
    \label{fig:ddo44}
\end{figure*}

\subsection{H\,{\sc i} detections}
Since the cube resulting from the combination of the new and archival data has a lower noise, we use it to create the detection mask by running the H\,{\sc i} Source Finding Application (SoFiA2) v2.4.0 \citep{sofia,sofia2}. SoFiA2 internally takes care of the uneven noise, because it uses the global noise as the flux threshold for detecting emission. We blanked the channels from $-55$ km\,s$^{-1}$ to 5 km\,s$^{-1}$ since they contain strong Milky Way contamination, and used SoFiA2 to identify the emission in the data cube using the S+C method. We used a spatial kernel equal to 0, 1 and 2 times the beam size (i.e., a kernel of $0\times0$, $3.1\times3.1$ and $6.2\times6.2$ pixels) and a spectral kernel equal to 0, 3 and 7 channels, a detection threshold of 3$\sigma$ and a reliability threshold of 0.95.\\We produced also the moment 0, 1 and 2 maps by setting the linker in SoFiA2 to be a $1\times1\times1$ box ($1\times1$ pixel and 1 channel) and by rejecting all the sources whose size is less than $6\times6\times3$ ($6\times6$ pixel and 3 channels). The moment maps are given in Fig. \ref{fig:maps}. We calculated an H\,{\sc i} mass for the galaxy of 3.2 $\times$ 10\textsuperscript{9} M$_{\odot}$, which agrees with previous radio interferometric measurements \citep{fraternali02,things}.\\In Fig. \ref{fig:chanmap} we show the channel maps covering the velocity range over which H\,{\sc i} emission is found. From now on, all the velocities are referred to the systemic velocity of 133.2 km\,s$^{-1}$. Apart from the well-known 8-kpc filament \citep{fraternali02,blok14}, we indicate three additional filamentary structures, pointed out in Fig. \ref{fig:chanmap} with the arrows:
\begin{itemize}
    \item between $-$59 km\,s$^{-1}$ and $-$27 km\,s$^{-1}$ on the east side (red arrow in Fig. \ref{fig:chanmap});
    \item from $-$27 km\,s$^{-1}$ to 15 km\,s$^{-1}$ on the west side of the galaxy  (blue arrow in Fig. \ref{fig:chanmap}) next to the 8-kpc filament;
    \item from 15 km\,s$^{-1}$ to 47 km\,s$^{-1}$ on the west side of the galaxy  (green arrow in Fig. \ref{fig:chanmap}) above the 8-kpc filament.
\end{itemize}
The 8-kpc filament found in previous work may be part of a more extended H\,{\sc i} stream as clearly shown in the 15 km\,s$^{-1}$ channel, whose total extent is at least 20 kpc. We will denote this as the `20 kpc filament' or `main filament'. The high column density parts of the other two structures were detected already in \citet{fraternali01}, but the increased column density sensitivity of the new data presented here highlights their filamentary shape and extent.\\The location of those features in velocity space implies that they are not co-rotating with the thin disk. It is thus possible to kinematically isolate them from the thin disk emission for a detailed study.

\subsection{Extraction of the H\,{\sc i} filaments}
As a first step, we isolate the thin disk following the method presented in \citet{fraternali02}. We fit a single Gaussian to the line profile for each position in the sky resulting in a `Gaussian cube' that we subtracted from the data cube.\\We then used the velocities of the peak values of these fitted profiles to line up all H\,{\sc i} emission profiles for each position on the sky at their peak velocities. This method is called `shuffle' and was applied to both the cube and the SoFiA2 detection mask produced before. It is a powerful tool to easily isolate the anomalous velocity gas, for example as a function of the deviation from the thin disk rotation as shown in Fig. \ref{fig:shufflepv}. Here, the gas rotating with the disk occupies the central channels of the shuffled cube, while H\,{\sc i} that is not rotating with the disk will be placed at higher/lower channels. Hence, the non-co-rotating gas can be easily identified, as we show in Fig. \ref{fig:shufflepv}.\\We isolated the channels of the shuffled cube where the filamentary emission is present by blanking the channels between $\pm$ 65 km\,s$^{-1}$ (or equivalently $\pm$ 14 times the full width at half maximum of the disk profiles). This selection is conservative by design. Our purpose is to isolate the filaments with the highest velocities that are most likely not associated with the thin disk or the thicker and lagging disk as identified in \citet{fraternali02}. A narrower velocity cut than used here would include a significant amount of gas from this lagging, thick disk. We calculated the moment maps of the anomalous-velocity gas using the shuffled detection mask. The result is given in Fig. \ref{fig:filmap}, where we show the overlay between the galaxy emission and the GBT detection (left panel), the filamentary complex (central panel) and the selection effect introduced by the shuffle procedure (right panel), which we will discuss in the next section.

\section{Discussion}\label{sec:disk}
The H\,{\sc i} filaments have a total mass of at least 2.0 $\times$ 10\textsuperscript{7} M$_{\odot}$, about 0.6\% of the total H\,{\sc i} mass of the galaxy. This is a lower limit, firstly because of the chosen channel cut at $\pm$ 65 km\,s$^{-1}$, and secondly because close to the kinematical minor axis of the galaxy, any anomalous velocity gas co-rotating or counter-rotating with the disk will not be distinguishable. In other words, as we look close to the kinematical minor axis, the line profile of the thin disk and of the extraplanar gas overlap more and more, and we are not able to distinguish between thin disk and co-rotating or counter-rotating anomalous velocity gas.\\We used the following simple model to estimate the importance of this effect: we built a mock galaxy with a thin (dispersion of 10 km\,s$^{-1}$) thin disk and a thick (dispersion of 35 km\,s$^{-1}$) disk which rotates 30 km\,s$^{-1}$ slower and has a column density equal to 10\% of the thin disk. These parameters are modelled on the properties of NGC 2403 as described in \citet{fraternali02}.\\We repeated the same steps applied to the observed data (Gaussian fit, thin disk removal, shuffle and blanking of all the channels between $\pm$ 65 km\,s$^{-1}$) in order to check how our ability to extract the anomalous H\,{\sc i} varies as a function of the position angle along the disk. We do indeed find that the ability to separate the anomalous velocity gas from the thin disk changes with position angle. In the right panel of Fig. \ref{fig:filmap} we illustrate the effect.\\The projected extent of the filaments varies from $\sim$ 10 kpc up to $\sim$ 20 kpc, and they are almost aligned with the galactic major axis. This is not purely a result of selection effects, as the distribution of the observed anomalous H\,{\sc i} in the right panel of Fig. \ref{fig:filmap} clearly shows.\\The GBT cloud found by \citet{blok14} is at the location of the tip of the 20 kpc filament and can possibly be identified with it, resembling a cloud in the GBT data only due to the very low spatial resolution of these data and the limited velocity range over which this feature could be identified.\\The larger extension of the main filament as visible in the new data due to the increased sensitivity is shown in Fig. \ref{fig:filpv}, where we show the position-velocity diagram along the main filament and the cloud compared to the same position velocity slice presented in Fig. 6 in \citet{blok14}.\\In the following we discuss further the possible origins of this filamentary complex.

\subsection{Is the GBT cloud confirmed?}
The motivation of this study was to confirm the GBT detection of an H\,{\sc i} cloud near NGC 2403 \citep{blok14} and to understand its origin. The left panel of Fig. \ref{fig:filmap} shows that the GBT cloud overlaps with the northern part of the 20 kpc filament. Indeed, the column density of the gas in the northern region of the filament is $>$ 10\textsuperscript{18} cm\textsuperscript{-2} at the spatial resolution of $\sim$ 50 arcsec and \citet{blok14} estimates an average column density of 2.4 $\times$ 10\textsuperscript{18} cm\textsuperscript{-2} for the cloud. Therefore, the tip of the 20 kpc filament could have been seen by the GBT.\\The H\,{\sc i} mass of the cloud as measured with the GBT is 6.3 $\times$ 10\textsuperscript{6} M$_{\odot}$, while the H\,{\sc i} content within the GBT contours in our data is 1.2 $\times$ 10\textsuperscript{6} M$_{\odot}$. Given the inherent difficulties of separating the cloud and disk emission in the GBT data and the conservative blanking criteria used to extract the filamentary emission in our shuffle cube, this must be regarded as reasonable agreement and thus further supports the identification of the GBT cloud with the northern tip of the filament detected by the VLA.

\subsection{The nature of the 20 kpc filament}
One of the questions we try to answer is whether the GBT cloud, and by implication the filaments described in this paper, are formed due to a gas accretion event. There are however other possible explanations for these features, some of which we have already briefly alluded to. For example, the structures we observe could be the result of a galactic fountain process that ejects gas from the disk. This is certainly a likely scenario for the anomalous velocity gas closer to the disk (i.e. below our $\pm$ 65 km\,s$^{-1}$ cut-off), but it is less obvious for the filaments.\\It is difficult to form a straight 20 kpc long filament in an outflow scenario. Indeed, in a pure galactic fountain model the gas ejecta reaches distances of at most 10 kpc even with extremely high kick velocities \citep{fraternali06,fraternali08}.\\Is the origin of the filamentary complex a galactic interaction? In the event of a merging or disrupting dwarf, one might expect to see a disturbed stellar population from the dwarf, most likely coinciding with or near the filaments. \citet{barker12} performed an in-depth study of the stellar properties in NGC 2403 but did not identify any peculiarities in the distribution of the stars and concluded that the galaxy has evolved in isolation with no significant recent merging event.\\However, their Fig. 10 shows some evidence for a slightly disturbed stellar component in the NW part of the disk. They also reported an extended low-metallicity Red Giant Branch (RGB) component at radii between 18 and 40 kpc. Their interpretation includes the presence of a stellar halo, or a thick stellar disk in NGC 2403, but it is not clear whether this component is associated with an accretion or interaction event.\\In \citet{blok14} it was shown that a line passing through the major axis of the 8-kpc filament and the GBT cloud intersects with nearby dwarf galaxies DDO 44 and NGC 2366, hinting at a possible interaction scenario with either of these galaxies. Recent deep optical wide-field data presented in \citet{carlin19} supports this idea. They observed a $\sim$ 50 kpc long stellar stream originating from the dwarf galaxy DDO 44 and connecting with the NW part of the stellar disk of NGC 2403. This dwarf is likely a satellite of NGC 2403 located about 70 kpc from it. A higher quality re-reduction of the data from \citet{carlin19} is given in Fig. \ref{fig:ddo44} (J. Carlin, priv. comm) and this shows the stellar stream and the connection between DDO 44 and NGC 2403 even more clearly.\\This image is a 1.5$^{\circ}$ wide field composite mosaic of seven pointings taken with the Hyper Suprime-Cam of the 8.2 m Subaru Telescope and converted into a density map of candidate RGB stars. Overlaid on that map, we show the filamentary H\,{\sc i}. There is a remarkable coincidence between the position of the tip of the 20 kpc filament and the region where the DDO 44 stellar stream connects with NGC 2403 stellar disk. This is therefore a strong indication that the dwarf interaction scenario can be responsible for the H\,{\sc i} filaments in NGC 2403.\\This is also suggested by the star formation histories of the galaxies: NGC 2403 had an increase in its star formation rate about 2 Gyr ago \citep{williams13}, in particular, at radii $>$ 20 kpc; while DDO 44 had a burst of star formation from 1 to 3 Gyr ago \citep{girardi10,weisz11}. Also, \citet{carlin19} calculated from the estimated trajectory of DDO 44 that the interaction with NGC 2403 occurred $\sim$ 1 Gyr ago. These various estimates thus give very similar timescales for the interaction.\\The most likely explanation is therefore that the H\,{\sc i} filaments in NGC 2403 are the result of the interaction with DDO 44. This does, of course, not preclude a galactic fountain scenario as a potential origin for a part of the extraplanar gas in NGC 2403 especially so for the gas at less extreme velocities than the filaments. It is interesting to note that for two of the most well-known nearby galaxies where extraplanar filaments are prominent, NGC 2403 and NGC 891, an interaction scenario is now the most likely cause (see \citealt{oosterloo07} for the NGC 891 analysis).

\section{Conclusion}\label{sec:conc}
Using new and archival VLA observations, we have presented a study of NGC 2403, focusing on the nature of the GBT cloud \citep{blok14}. These data show a filamentary complex around the galaxy with an H\,{\sc i} mass of 2.0 $\times$ 10\textsuperscript{7} M$_{\odot}$. It comprises at least three filaments with extents of 10-20 kpc. The GBT cloud can now be shown to be the tip of the longest structure.\\Even though at first sight the stellar distribution in NGC 2403 seems undisturbed, it is now likely that these filaments are the result of an interaction with a nearby dwarf galaxy. Hints of a disturbance in the stellar component were already visible in the data presented in \citet{barker12}, but they were dramatically confirmed in the recent study by \citet{carlin19}, who discovered a $\sim$ 50 kpc long stellar stream connecting DDO 44 and NGC 2403. The intersection of that stream with NGC 2403 coincides with the positions of the tip of the H\,{\sc i} filaments. The star formation history of the two galaxies and the estimated trajectory of DDO 44 seems to suggest that the interaction occurred about 1 Gyr ago.\\These observations highlight the importance of minor interactions in shaping the H\,{\sc i} disks of galaxies. Moreover, they point out the difficulties of unambiguously identifying the effects of direct cold gas accretion from the IGM, and distinguishing them from those of minor interactions with a satellite. In this regard, the multiwavelength approach is fundamental: optical observations are crucial to move the needle toward accretion or interaction as the most plausible explanation of anomalous H\,{\sc i} structures like the filaments.\\Future observations with the SKA and its precursors at higher resolution and higher sensitivity will undoubtedly uncover many more of these features in nearby galaxies and help us to better understand how the balance between accretion and interaction impacts the evolution of galaxies.

\begin{acknowledgements}
We would like to thank A. Fergusson for providing us with the stellar catalogue from Barker et al. 2012. We thank F. Fraternali for the useful discussions about the interpretation of our results and F. Maccagni for the initial discussions on the DDO 44 optical data. A. Marasco's help in the modelling of our data is gratefully acknowledged. Also, we thank J. Carlin for kindly sharing with us the new re-reduced optical data of DDO 44. We thank the anonymous referee for the constructive comments and the suggested improvements for the quality and clarity of this paper.\\This work has received funding from the European Research Council (ERC) under the European Union’s Horizon 2020 research and innovation programme (grant agreement No 882793 `MeerGas').
\end{acknowledgements}

\bibliographystyle{aa}
\bibliography{reference.bib}

\end{document}